   \documentstyle[12pt]{article}
\if@twoside\oddsidemargin25mm
\else\oddsidemargin25mm\evensidemargin25mm\marginparwidth25mm\fi%
\begin{document}
\newcommand{\ft}[2]{{\textstyle\frac{#1}{#2}}}
\newcommand{\QED}{{\hspace*{\fill}\rule{2mm}{2mm}\linebreak}}
\def\dop{{\rm d}\hskip -1pt}
\def\bfone{\relax{\rm 1\kern-.35em 1}}
\def\bfzero{\relax{\rm I\kern-.18em 0}}
\def\inbar{\vrule height1.5ex width.4pt depth0pt}
\def\IC{\relax\,\hbox{$\inbar\kern-.3em{\rm C}$}}
\def\ID{\relax{\rm I\kern-.18em D}}
\def\IF{\relax{\rm I\kern-.18em F}}
\def\IK{\relax{\rm I\kern-.18em K}}
\def\IH{\relax{\rm I\kern-.18em H}}
\def\II{\relax{\rm I\kern-.17em I}}
\def\IN{\relax{\rm I\kern-.18em N}}
\def\IP{\relax{\rm I\kern-.18em P}}
\def\IQ{\relax\,\hbox{$\inbar\kern-.3em{\rm Q}$}}
\def\IR{\relax{\rm I\kern-.18em R}}
\def\IG{\relax\,\hbox{$\inbar\kern-.3em{\rm G}$}}
\font\cmss=cmss10 \font\cmsss=cmss10 at 7pt
\def\ZZ{\relax\ifmmode\mathchoice
{\hbox{\cmss Z\kern-.4em Z}}{\hbox{\cmss Z\kern-.4em Z}}
{\lower.9pt\hbox{\cmsss Z\kern-.4em Z}}
{\lower1.2pt\hbox{\cmsss Z\kern-.4em Z}}\else{\cmss Z\kern-.4em
Z}\fi}
\def\a{\alpha} \def\b{\beta} \def\d{\delta}
\def\e{\epsilon} \def\c{\gamma}
\def\G{\Gamma} \def\l{\lambda}
\def\L{\Lambda} \def\s{\sigma}
\def\cA{{\cal A}} \def\cB{{\cal B}}
\def\cC{{\cal C}} \def\cD{{\cal D}}
\def\cF{{\cal F}} \def\cG{{\cal G}}
\def\cH{{\cal H}} \def\cI{{\cal I}}
\def\cJ{{\cal J}} \def\cK{{\cal K}}
\def\cL{{\cal L}} \def\cM{{\cal M}}
\def\cN{{\cal N}} \def\cO{{\cal O}}
\def\cP{{\cal P}} \def\cQ{{\cal Q}}
\def\cR{{\cal R}} \def\cV{{\cal V}}\def\cW{{\cal W}}
%
%
%
\def\crr{\crcr\noalign{\vskip {8.3333pt}}}
\def\tilde{\widetilde}
\def\bar{\overline}
\def\us#1{\underline{#1}}
\let\shat=\hat
\def\hat{\widehat}
\def\hyp{\vrule height 2.3pt width 2.5pt depth -1.5pt}
\def\square{\mbox{.08}{.08}}
\def\Coeff#1#2{{#1\over #2}}
\def\Coe#1.#2.{{#1\over #2}}
\def\coeff#1#2{\relax{\textstyle {#1 \over #2}}\displaystyle}
\def\coe#1.#2.{\relax{\textstyle {#1 \over #2}}\displaystyle}
\def\half{{1 \over 2}}
\def\shalf{\relax{\textstyle {1 \over 2}}\displaystyle}
\def\dag#1{#1\!\!\!/\,\,\,}
\def\to{\rightarrow}
\def\notin{\hbox{{$\in$}\kern-.51em\hbox{/}}}
\def\shdot{\!\cdot\!}
\def\ket#1{\,\big|\,#1\,\big>\,}
\def\bra#1{\,\big<\,#1\,\big|\,}
\def\equaltop#1{\mathrel{\mathop=^{#1}}}
\def\Trbel#1{\mathop{{\rm Tr}}_{#1}}
\def\inserteq#1{\noalign{\vskip-.2truecm\hbox{#1\hfil}
\vskip-.2cm}}
\def\attac#1{\Bigl\vert
{\phantom{X}\atop{{\rm\scriptstyle #1}}\phantom{X}}}
\def\exx#1{e^{{\displaystyle #1}}}
\def\del{\partial}
\def\delbar{\bar\partial}
\def\nex#1{$N\!=\!#1$}
\def\dex#1{$d\!=\!#1$}
\def\cex#1{$c\!=\!#1$}
\def\eg{{\it e.g.}} \def\ie{{\it i.e.}}
%
\def\cS{{\cal K}}
\def\IE{\relax{{\rm I\kern-.18em E}}}
\def\cE{{\cal E}}
\def\rt{{\cR^{(3)}}}
\def\IGam{\relax{{\rm I}\kern-.18em \Gamma}}
\def\IGa{\IA}
\def\LG{Lan\-dau-Ginz\-burg\ }
\def\cV{{\cal V}}
\def\Rt{{\cal R}^{(3)}}
\def\wabc{W_{abc}}
\def\WABC{W_{\a\b\c}}
\def\W{{\cal W}}
\def\tft#1{\langle\langle\,#1\,\rangle\rangle}
\def\IA{\relax{\hbox{{\rm A}\kern-.82em {\rm A}}}}
\let\picfuc=\fp
\def\hata{{\shat\a}}
\def\hatb{{\shat\b}}
\def\hatA{{\shat A}}
\def\hatB{{\shat B}}
\def\bv{{\bf V}}
\def\spg{special geometry}
\def\sc{SCFT}
\def\leel{low energy effective Lagrangian}
\def\pf{Picard--Fuchs}
\def\pfS{Picard--Fuchs system}
\def\el{effective Lagrangian}
\def\Fb{\overline{F}}
\def\nablab{\overline{\nabla}}
\def\Ub{\overline{U}}
\def\Db{\overline{D}}
\def\zb{\overline{z}}
\def\eb{\overline{e}}
\def\fb{\overline{f}}
\def\tb{\overline{t}}
\def\Xb{\overline{X}}
\def\Vb{\overline{V}}
\def\Cb{\overline{C}}
\def\Sb{\overline{S}}
\def\delb{\overline{\del}}
\def\Gammab{\overline{\Gamma}}
\def\Ab{\overline{A}}
\def\Anh{A^{\rm nh}}
\def\alphab{\bar{\alpha}}
\def\cy{Calabi--Yau}
\def\cabg{C_{\alpha\beta\gamma}}
\def\B{\Sigma}
\def\Bh{\hat \Sigma}
\def\Kh{\hat{K}}
\def\Knh{{\cal K}}
\def\A{\Lambda}
\def\Ah{\hat \Lambda}
\def\R{\hat{R}}
\def\V{{V}}
\def\T{T}
\def\Gammah{\hat{\Gamma}}
\def\twot{$(2,2)$}
\def\K{K\"ahler}
\def\rat{({\theta_2 \over \theta_1})}
\def\lv{{\bf \omega}}
\def\w{w}
\def\CP{C\!P}
\def\o#1#2{{{#1}\over{#2}}}
\newcommand{\be}{\begin{equation}}
\newcommand{\ee}{\end{equation}}
\newcommand{\ba}{\begin{eqnarray}}
\newcommand{\ea}{\end{eqnarray}}
\newtheorem{definizione}{Definition}[section]
\newcommand{\bd}{\begin{definizione}}
\newcommand{\ed}{\end{definizione}}
\newtheorem{teorema}{Theorem}[section]
\newcommand{\bth}{\begin{teorema}}
\newcommand{\eth}{\end{teorema}}
\newtheorem{lemma}{Lemma}[section]
\newcommand{\blem}{\begin{lemma}}
\newcommand{\elem}{\end{lemma}}
\newcommand{\brr}{\begin{array}}
\newcommand{\err}{\end{array}}
\newcommand{\nn}{\nonumber}
\newtheorem{corollario}{Corollary}[section]
\newcommand{\bcorol}{\begin{corollario}}
\newcommand{\ecorol}{\end{corollario}}
\def\twomat#1#2#3#4{\left(\begin{array}{cc}
 {#1}&{#2}\\ {#3}&{#4}\\
\end{array}
\right)}
\def\twovec#1#2{\left(\begin{array}{c}
{#1}\\ {#2}\\
\end{array}
\right)}
\begin{titlepage}
\hskip 12cm
\vbox{\hbox{hep-th/9703156}\hbox{March, 1997}}
\vfill
\begin{center}
{\LARGE { U-Invariants, Black-Hole Entropy and Fixed Scalars}}\\
\vskip 1.5cm
{  {\bf Laura Andrianopoli$^1$,
Riccardo D'Auria$^2$ and
Sergio Ferrara$^3$ }} \\
\vskip 0.5cm
{\small
$^1$ Dipartimento di Fisica, Universit\'a di Genova, via Dodecaneso 33,
I-16146 Genova\\
and Istituto Nazionale di Fisica Nucleare (INFN) - Sezione di Torino, Italy\\
\vspace{6pt}
$^2$ Dipartimento di Fisica, Politecnico di Torino,\\
 Corso Duca degli Abruzzi 24, I-10129 Torino\\
and Istituto Nazionale di Fisica Nucleare (INFN) - Sezione di Torino, Italy\\
\vspace{6pt}
$^3$ CERN Theoretical Division, CH 1211 Geneva 23, Switzerland
}
\end{center}
\vfill
\begin{center} {\bf Abstract}
\end{center}
{
\small
The  absolute  (moduli-independent)  U-invariants  of  all  $N>2$ extended
supergravities   at  $D=4$  are  derived  in  terms  of  (moduli-dependent)
central and matter charges.
\\
These  invariants  give a general definition of the ``topological''
Bekenstein-Hawking  entropy  formula for extremal black-holes and reduce to
the  square  of the black-hole ADM mass for ``fixed scalars'' which extremize
the black-hole ``potential'' energy.
\\
The  Hessian matrix of the black-hole potential 
 at  ``fixed  scalars'', in contrast to $N=2$ theories, is shown to
be  degenerate,  with  rank  $(N-2)(N-3) + 2 n$ ($N$ being the number
 of supersymmetries
and $n$ the number of matter multiplets) and  semipositive definite.
}
\vspace{2mm} \vfill \hrule width 3.cm
{\footnotesize
\noindent
$^*$ Work supported in part by EEC under TMR contract ERBFMRX-CT96-0045
 (LNF Frascati, 
Politecnico di Torino and Univ. Genova) and by DOE grant
DE-FGO3-91ER40662}
\end{titlepage}
\section{Introduction}
Recently, considerable progress has been made in the study of general 
properties 
of black holes arising in supersymmetric theories of gravity such as  extended
supergravities, string theory and M-theory \cite{string}.
Of particular interest are extremal black holes in four dimensions which 
correspond to BPS saturated states \cite{black} and whose ADM mass depends,
 beyond the
quantized values of electric and magnetic charges, on the asymptotic value
 of scalars at infinity.
The latter describe the moduli space of the theory.

Another physical relevant quantity, which depends only on quantized electric
 and magnetic charges,
 is the black hole entropy,
which can be defined macroscopically, through the Bekenstein-Hawking
 area-entropy relation
or microscopically, through D-branes techniques \cite{dbr} by counting 
of microstates \cite{micros}.
It has been further realized that the scalar fields, independently of 
their values
at infinity, flow towards the black hole horizon to a fixed value of pure
 topological
nature given by a certain ratio of electric and magnetic charges \cite{fks}.

These ``fixed scalars'' correspond to the extrema of the ADM mass
in moduli space while the black-hole entropy  is the actual value of the
 squared
ADM mass at this point \cite{feka1}.

In theories with $N>2$, extremal black-holes preserving one supersymmetry 
have
the further property that all central charge eigenvalues other than the one
equal to the BPS mass flow to zero for ``fixed scalars''.
The black-hole entropy is still given by the square of the ADM mass for
 ``fixed scalars''\cite{feka2}. 

Recently \cite{fegika}, the nature of these extrema has been further studied 
and shown that  they
generically correspond to non degenerate minima for $N=2$ theories whose
 relevant
moduli space is the special geometry of $N=2$ vector multiplets.

The entropy formula turns out to be in all cases a U-duality invariant 
expression
(homogeneous of degree two) built out of electric and magnetic charges and as
 such
can be in fact also computed through certain (moduli-independent) topological 
quantities which only depend on the nature of the U-duality groups and the 
appropriate representations
of electric and magnetic charges.
For example, in the $N=8$ theory  the entropy was shown to correspond  to the
unique quartic $E_7$ invariant built with its 56 dimensional representation 
\cite{kall}.

In this paper we 
intend to make further progress in this subject by  deriving, for all $N>2$ 
theories, 
topological (moduli-independent) U-invariants constructed in terms of 
(moduli-dependent)
 central charges and matter charges, and show that, as expected, they coincide 
with the squared ADM mass at ``fixed scalars''.

We also show that the Hessian  of the  black-hole potential, as it arises
from the geodesic action \cite{bmgk}\cite{fegika}, is degenerate at the 
extremum
 for $N>2$ theories, and we discuss the nature of this degeneracy.
The Hessian is of physical relevance because it is related to a
 thermodynamical quantity named Weinhold metric \cite{fegika}.

The paper is organized as follows:
\\
In section 2 we give the topological U-invariants for all $N>2$ supergravities
 and show,
in an appendix, how to derive them with a simple mathematical tool.
In section 3 we discuss the degeneracy and the rank of the Hessian matrix.
In section 4 we discuss these results in a string theory perspective.

\section{Central charges, U-invariants and entropy}
Extremal   black-holes   preserving   one   supersymmetry  correspond  to
$N$-extended multiplets with
\begin{equation}
M_{ADM} = \vert Z_1 \vert >  \vert Z_2 \vert  \cdots > \vert Z_{[N/2]} \vert
         \end{equation}
         where $Z_\alpha$, $\alpha =1,\cdots, [N/2]$, are the proper values of
 the
         central charge antisymmetric matrix written in normal form
 \cite{fesazu}.
The central charges $Z_{AB}= -Z_{BA}$, $A,B=1,\cdots,N$, and matter charges
 $Z_I$, $I= 1,\cdots , n$ are
those (moduli-dependent) symplectic invariant combinations of field strenghts 
and their duals
(integrated over a large two-sphere)
 which appear
in the gravitino and gaugino supersymmetry variations respectively 
\cite{cedafe}, \cite{noi1}, \cite{noi}.
Note that the total number of vector fields is $n_v=N(N-1)/2+n$ (with the
 exception of $N=6$ 
in which case there is an extra singlet graviphoton)\cite{cj}.
         \\
  It was shown in ref. \cite{feka2} that at the attractor point, where
  $M_{ADM}$ is extremized, supersymmetry requires that $Z_\alpha$, $\alpha >1$,
  vanish toghether with the matter charges $Z_I$, $I= 1, \cdots , n$
($n$ is the number of matter multiplets, which can exist only for $N=3,4$)
\par
This result can be used to show that for ``fixed scalars'',
corresponding to the attractor point, the scalar ``potential'' of the geodesic
 action \cite{bmgk}\cite{fegika}
\begin{equation}
V=-{1\over 2}P^t\cM(\cN)P 
\label{sumrule1}
\end{equation}
is extremized in moduli space.
Here $P$ is the symplectic vector $P=(p^\Lambda, q_\Lambda) $ of quantized 
electric and
magnetic charges and $\cM(\cN)$ is a symplectic $2n_v \times 2n_v$ matrix
 whose 
$n_v\times n_v$ blocks are given in terms of the $n_v\times n_v$ vector
 kinetic matrix $\cN_{\Lambda\Sigma}$
($-Im \cN, Re \cN$ are the normalizations of the kinetic $F^2$ and the 
topological $F^*F$
terms respectively) and 
\begin{equation}
\cM(\cN) = \pmatrix{A& B \cr C & D \cr}
\end{equation}
with:
\begin{eqnarray}
A&=& Im \cN + Re \cN Im \cN^{-1} Re\cN \nonumber\\
B&=& - Re \cN Im \cN^{-1} \nonumber\\
C&=& -  Im \cN^{-1} Re\cN \nonumber\\
D&=&Im \cN^{-1} 
\end{eqnarray}
The above assertion comes from the important identity, shown in ref.
 \cite{noi1}, \cite{noi} to be valid 
in all $N\geq2$ theories:
\begin{equation}
-{1\over 2}P^t\cM(\cN)P = {1\over 2} Z_{AB} \bar Z^{AB} + Z_I \bar Z^I
\end{equation}
The main purpose of this section is to provide  particular expressions which
 give the
entropy formula as a moduli--independent quantity in the entire
moduli space and not just at the critical points.
Namely, we are looking for quantities $S\left(Z_{AB}(\phi), \bar Z^{AB}
 (\phi),Z_{I}(\phi), \bar Z^{I} (\phi)\right)$
such that ${\partial \over \partial \phi ^i} S =0$, $\phi ^i$ being the moduli
 coordinates
 \footnote{The Bekenstein-Hawking entropy $S_{BH} ={A\over 4}$ is actually 
$\pi S$
in our notation.}.
 
These formulae generalize the quartic $E_{7(-7)}$ invariant of $N=8$
supergravity \cite{kall} to all other cases. 
We will show in the appendix how these invariants can be computed in an almost
trivial fashion by using the (non compact) Cartan elements of $G/H$.
\par
Let us first consider the theories $N=3,4$, where  matter can be
present \cite{maina}, \cite{bks}.
\par
The U--duality groups \footnote{Here we denote by U-duality group the isometry
 group $G$
acting on the scalars, although only a restriction of it to integers is the
 proper U-duality group \cite{ht}.}
 are, in these cases, $SU(3,n)$ and $SU(1,1)
\times SO(6,n)$ respectively.
The central and matter charges $Z_{AB}, Z_I$ transform in an obvious
way under the isotropy groups
\begin{eqnarray}
H&=& SU(3) \times SU(n) \times U(1) \qquad (N=3) \\
 H&=& SU(4) \times O(n) \times U(1) \qquad (N=4)
\end{eqnarray}
Under the action of the elements of $G/H$ the charges get mixed with
their complex conjugate. The infinitesimal transformation
 can be read from the differential relations satisfied by the charges
 \cite{noi}:
 \begin{eqnarray}
   \nabla Z_{AB} &=& {1\over 2} P_{ABCD} \bar Z^{CD} + P_{AB I} \bar Z^I \\
 \nabla Z_{I} &=& {1\over 2} P_{AB I} \bar Z^{AB} + P_{IJ} \bar Z^J
 \end{eqnarray}
where the matrices $P_{ABCD}$, $P_{AB I}$, $P_{IJ}$ are the subblocks of the
 vielbein of $G/H$ \cite{noi}:
\begin{equation}
  \label{vielbein}
  \cP \equiv L^{-1} \nabla L = \pmatrix{P_{ABCD}& P_{AB I} \cr
P_{I AB} & P_{IJ} \cr }
\end{equation}
written in terms of the indices of $H=H_{Aut} \times H_{matter}$.
\par
For $N=3$:
\begin{equation}
P^{ABCD}=P_{IJ}=0 , \quad P_{ AB I} \equiv \epsilon_{ABC}P^C_I \quad Z_{AB}
 \equiv \epsilon_{ABC}Z^C
 \label{viel3} 
\end{equation}
Then  the variations are:
\begin{eqnarray}
\delta Z^A  &=& \xi^A _I \bar Z^I  \\
\delta Z_{I}  &=&\xi^A _{ I} \bar Z_A
\label{deltaz3}
\end{eqnarray}
where $\xi^A_I$ are infinitesimal parameters of $K=G/H$.
\par
So, the U-invariant expression is:
\begin{equation}
S=   Z^A \bar Z_A - Z_I \bar Z^I
\label{invar3}
\end{equation}
In other words, $\nabla_i S = \partial_i S =0 $, where the covariant 
derivative is
defined in ref. \cite{noi}.
\par
Note that at the attractor point ($Z_I =0$) it coincides with the
moduli-dependent potential (\ref{sumrule1}) 
computed at its extremum.
\par
For $N=4$
\begin{equation}
P_{ABCD} = \epsilon_{ABCD}P ,\quad P_{IJ} = \eta_{IJ}P,\quad
P_{AB I}={1\over 2} \eta_{IJ} \epsilon_{ABCD}\bar P^{CD J} \label{viel4}
\end{equation}
and the transformations of $K= {SU(1,1) \over U(1)} \times
{O(6,n) \over O(6) \times O(n)}$ are:
 \begin{eqnarray}
\delta Z_{AB}  &=& {1\over 2} \xi  \epsilon_{ABCD} \bar Z^{CD}  +
 \xi _{AB I} \bar Z^I \\
\delta Z_{I}  &=& \xi \eta_{IJ} \bar Z^J + {1\over 2}\xi _{AB I} \bar Z^{AB}
\label{deltaz4}
\end{eqnarray}
with $\bar \xi^{AB I} =  {1\over 2} \eta^{IJ}   \epsilon^{ABCD}
\xi_{CD J}$.
\par
There are three $O(6,n)$ invariants given by $I_1$, $I_2$, $\bar I_2$ where:
\begin{eqnarray}
  I_1 &=& {1 \over 2}  Z_{AB} \bar Z_{AB} - Z_I \bar Z^I
\label{invar41} \\
I_2 &=& {1\over 4} \epsilon^{ABCD}  Z_{AB}   Z_{CD} - \bar Z_I \bar Z^I
\label{invar42}
\end{eqnarray}
and the unique   $SU(1,1)  \times
O(6,n) $  invariant $S$, $\nabla S =0$, is given by:
\begin{equation}
S= \sqrt{(I_1)^2 - \vert I_2 \vert ^2 }
\label{invar4}
\end{equation}

At the attractor point $Z_I =0 $ and $\epsilon^{ABCD} Z_{AB} Z_{CD}
=0$ so that $S$ reduces to the square of the BPS mass.
\par
For $N=5,6,8$ the U-duality invariant expression $S$ is the square
root of a unique invariant under the corresponding U-duality groups
$SU(5,1)$, $O^*(12)$ and $E_{7(-7)}$.
The strategy is to find a quartic expression $S^2$    in terms of
$Z_{AB}$ such that $\nabla S=0$, i.e. $S$ is moduli-independent.
\par
As before, this quantity is a particular combination of the $H$
quartic invariants.
\par
For $SU(5,1)$ there are only two  $U(5)$ quartic invariants.
In terms of the matrix $A_A^{\ B} = Z_{AC} \bar Z^{CB}$ they are:
$(Tr A)^2$, $Tr(A^2)$, where
\begin{eqnarray}
 Tr A & = & Z_{AB} \bar Z^{BA} \\
 Tr (A^2) & = & Z_{AB} \bar Z^{BC} Z_{CD} \bar Z^{DA}
\end{eqnarray}
As before, the relative coefficient is fixed by the transformation
properties of $Z_{AB}$ under ${SU(5,1) \over U(5) } $ elements of
infinitesimal parameter $\xi^C$:
\begin{eqnarray}
  \delta Z_{AB} = {1\over 2} \xi^C \epsilon_{CABPQ} \bar Z^{PQ}
\end{eqnarray}
It then follows that the required invariant is:
\begin{equation}
S= {1\over 2} \sqrt{  4 Tr(A^2) - (Tr A)^2 }
\label{invar5}
\end{equation}

For $N=8$ the $SU(8)$ invariants are \footnote{The Pfaffian of an
 $(n\times n)$ ($n$ even) antisymmetric
matrix is defined as $Pf Z={1\over 2^n n!} \epsilon^{A_1 \cdots A_n}
 Z_{A_1A_2}\cdots Z_{A_{N-1}A_N}$, with the property:
$ \vert Pf Z \vert = \vert det Z \vert ^{1/2}$. }:
\begin{eqnarray}
I_1 &=& (Tr A) ^2 \\
I_2 &=& Tr (A^2) \\
I_3 &=& Pf \, Z 
 ={1\over 2^4 4!} \epsilon^{ABCDEFGH} Z_{AB} Z_{CD} Z_{EF} Z_{GH}
\end{eqnarray}
The ${E_{7(-7)} \over SU(8)}$ transformations are:
\begin{equation}
\delta Z_{AB} ={1\over 2} \xi_{ABCD} \bar Z^{CD}
\end{equation}
where $\xi_{ABCD}$ satisfies the reality constraint:
\begin{equation}
\xi_{ABCD} = {1 \over 24} \epsilon_{ABCDEFGH} \bar \xi^{EFGH}
\end{equation}
One finds the following $E_{7(-7)}$ invariant \cite{kall}:
\begin{equation}
S= {1\over 2} \sqrt{4 Tr (A^2) - ( Tr A)^2 + 32 Re (Pf \,
Z) }
\end{equation}

The $N=6$ case is the more complicated because under $U(6)$ the
left-handed spinor of $O^*(12)$ splits into:
\begin{equation}
32_L \to (15,1) + ( \bar {15}, -1) + (1, -3) + (1,3)
\end{equation}
The transformations of ${O^*(12) \over U(6)}$ are:
\begin{eqnarray}
\delta Z_{AB} &=& {1\over 4} \epsilon_{ABCDEF} \xi^{CD} \bar Z^{EF} +
\xi_{AB} \bar X \\
\delta X &=& {1 \over 2} \xi _{AB} \bar Z^{AB}
\end{eqnarray}
 where we denote by $X$ the $SU(6)$ singlet.

The quartic $U(6)$ invariants are:
\begin{eqnarray}
I_1&=& (Tr A)^2 \label{invar61}\\
I_2&=& Tr(A^2)\label{invar62} \\
I_3 &=& Re (Pf \, ZX)= {1\over 2^3 3!}
Re( \epsilon^{ABCDEF}Z_{AB}Z_{CD}Z_{EF}X)\label{invar63}\\
I_4 &=& (Tr A) X \bar X\label{invar64}\\
I_5&=& X^2 \bar X^2\label{invar65}
\end{eqnarray}

The unique $O^*(12)$ invariant is:
\begin{eqnarray}
S&=&{1\over 2} \sqrt{4 I_2 - I_1 + 32 I_3 +4I_4 + 4 I_5 }
\label{invar6}   \\
\nabla S &=& 0
\end{eqnarray}
Note that at the attractor point $Pf\,Z =0$, $X=0$ and $S$ reduces to
the square of the BPS mass.

\section{Extrema of the BPS mass and fixed scalars}
In this section we would like to extend the analysis of the extrema of the 
black-hole induced potential 
\begin{equation}
V= {1\over 2} Z_{AB}\bar Z^{AB} + Z_I \bar Z^ I
\end{equation}
which was performed in ref \cite{fegika} for the $N=2$ case to all $N>2$
 theories.

We recall that, in the case of $N=2$ special geometry with metric
 $g_{i\bar\jmath}$,
 at the fixed scalar critical point
$\partial _i V=0$ the Hessian matrix reduces to:
\begin{eqnarray}
( \nabla_i \nabla _{\bar\jmath} V)_{fixed}&=& 
( \partial_i \partial _{\bar\jmath} V)_{fixed} = 2g_{i\bar\jmath}V_{fixed}\\
( \nabla_i \nabla _j V)_{fixed}&=&0
\end{eqnarray}
The Hessian matrix is strictly  positive-definite if the
critical point is not at the singular point of the
vector multiplet moduli-space.
This matrix was related to the Weinhold metric earlier  introduced in 
the geometric
approach to thermodynamics and used for the study of critical phenomena 
\cite{fegika}.

For $N$-extended supersymmetry, a form of this matrix was also given and
 shown to be
equal to \footnote{Generically the indices $i,j$ refer to real coordinates, 
unless
the manifold is K\"ahlerian, in which case we use holomorphic coordinates and
 formula (\ref{hessian})
reduces to the hermitean $i\bar\jmath$ entries of the Hessian matrix.}:
\begin{equation}
V_{ij}= ( \partial_i \partial _j V)_{fixed} = Z_{CD} Z^{AB} ({1\over 2}
 P^{CDPQ}_{\ \ \ \ ,j}P_{ABPQ,i} +  
P^{CD}_{I,j} P^I_{AB,j}).
\label{hessian}
\end{equation}

It is our purpose to further investigate properties of the Weinhold metric
 for fixed scalars.

Let us first observe that the extremum conditions $\nabla_i V=0$, using the
 relation 
between the covariant derivatives of the central charges, reduce to the
 conditions:
\begin{equation}
\epsilon^{ABCDL_1 \cdots L_{N-4}}Z_{AB} Z_{CD} = 0,\quad Z_I=0
\label{fixedsc}
\end{equation}
These equations give the fixed scalars in terms of electric and magnetic 
charges
and also show that the topological invariants of the previous section reduce
 to the extremum of the square of the ADM mass since, when the above
 conditions are fulfilled, 
$(Tr A)^2 = 2Tr(A^2)$, where $A_A^{\ B} = Z_{AB} \bar Z^{BC}$.

On the other hand, when these conditions are fulfilled, it is easy to see 
that the Hessian matrix is degenerate.
To see this, it is sufficient to go, making an $H$ transformation, to the 
normal frame in which these conditions imply $Z_{12} \neq 0 $ with the other
 charges vanishing.
Then we have:
\begin{equation}
\partial_i \partial_j V \vert_{fixed} = 4 \vert Z_{12}\vert ^2
({1\over 2} P^{12ab} _j P_{12ab,i} +  P^{ 12 I }_{,j} P_{12 I,i}) ,\quad
a,b \neq 1,2
\label{normetr}
\end{equation}

To understand the pattern of degeneracy for all $N$, we observe that when only
 one central 
charge in not vanishing the theory effectively reduces to an $N=2$ theory. 
Then the actual degeneracy respects $N=2$ multiplicity of the scalars degrees 
of freedom in the
sense that the degenerate directions will correspond to the hypermultiplet
 content
of $N>2$ theories when decomposed with respect to $N=2$ supersymmetry.

Note that for $N=3$, $N=4$, where $P_{AB I} $ is present, the Hessian is
block diagonal.

For $N=3$, referring to eq. (\ref{viel3}), since the  scalar manifold is
K\"ahler, $P_{AB I}$ is a (1,0)-form while $P^{ AB I}= \bar P_{AB I}$
is a (0,1)-form.
\par
The scalars appearing in the
$N=2$ vector multiplet and hypermultiplet content of the vielbein
are $P_{3I}$ for the vector multiplets and $P_{aI}$  ($a=1,2$) for
the hypermultiplets.
 From equation (\ref{normetr}), which for the $N=3$ case reads
 \begin{equation}
  \partial_{\bar \jmath} \partial_i V \vert_{fixed} = 2 \vert Z_{12}\vert ^2
 P_{3I,\bar \jmath} P^{3I}_{,i}
\end{equation}
we see that the metric has $4n$ real directions corresponding to
$n$ hypermultiplets  which are degenerate.
\par
For $N=4$, referring to (\ref{viel4}),
 $P$ is the $SU(1,1)/U(1)$
vielbein which gives one matter vector multiplet scalar while $P_{ 12 I}$ gives
$n$ matter vector multiplets.
The directions which are hypermultiplets correspond to $P_{ 1a I}, P_{ 2a I}$
 ($
a=3,4$).
Therefore the ``metric'' $V_{i j}$ is of rank $2n+2$.
\vskip 5mm
\par
For $N>4$, all the scalars are in the gravity multiplet and correspond to 
 $P_{ABCD}$.
\par
The splitting in vector and hypermultiplet scalars proceeds as
before.
Namely, in the $N=5$ case we set $P_{ABCD}=\epsilon_{ABCDL}P^L $
($A,B,C,D,L =1,\cdots 5$). In this case the vector multiplet scalars
are $P^a$ ($a=3,4,5$) while the hypermultiplet scalars are $P^1,
P^2$ ($n_V = 3$, $n_h =1$).
\par
For $N=6$, we set $P_{ABCD} ={1\over 2} \epsilon_{ABCDEF} P^{EF}$. The vector
multiplet scalars are now described by $P^{12}, P^{ab} $
($A,B,...=1,...,6$; $a,b = 3,\cdots 6$), while the hypermultiplet
scalars are given in terms of $P^{1a}, P^{2a} $.
Therefore we get $n_V = 6+1 =7$, $n_h =4$.
\par
This case is different from the others because, besides the
hypermultiplets $P^{1a}, P^{2a}$, also the vector multiplet direction
$P^{12}$ is degenerate.

Finally, for $N=8$ we have $P_{1abc}, P_{2abc}  $ as hypermultiplet
scalars and $P_{abcd}$ as vector multiplet scalars, which give
$n_V=15$, $n_h = 10$ (note that in this case the vielbein satisfies a
reality condition: $P_{ABCD}= {1\over 4! } \epsilon_{ABCDPQRS} \bar P
^{PQRS}$).
We have in this case 40  degenerate directions.
\par
In conclusion we see that the rank of the matrix $V_{ij} $ is $(N-2)
(N-3) + 2 n$ for all the four dimensional theories.

\section{Relations to string theories}
$N$-extended supergravities are related to strings compactified on
six-manifolds $M_N$ preserving $N$ supersymmetries at $D=4$.
Since we are presently considering $N>2$, the most common cases are
$N=4$ and $N=8$.
The first can be achieved in heterotic or Type II string, with $M_4 =
T_6$ in heterotic and $M_4 = K_3 \times T_2$ in Type II theory.
These theories are known to be dual at a non perturbative level \cite{ht},
 \cite{witten}, \cite{dlr}.
$N=8$ corresponds to $M_8 = T_6$ in Type II.
\par
Less familiar are the $N=3,5$ and 6 cases which were studied in ref.
\cite{fkff}.
\par
Interestingly enough, the latter cases can  be obtained by
compactification of Type II on asymmetric orbifolds with $3= 2_L +
1_R$, $5 = 4_L + 1_R$ and $6 = 4_L + 2_R$ respectively.
\par
BPS states considered in this paper should correspond to massive
states in these theories for which only a subset of them is known in
the perturbative framework.
\par
In attemps to test non perturbative string properties it would be
interesting to check the existence of the BPS states and their
entropy by using microscopic considerations.
\par
We finally  observe that, unlike $N=8$, the moduli spaces of
$N=3,5,6$ theories are locally K\"ahlerian (as $N=2$) with coset
spaces of rank 3 ($n \geq 3$), 1 and 3 respectively.
\par
For $N=5,6$ these spaces are also special  K\"ahler (which is also
the case for $N=3$ when $n = 1,3$) \cite{crvp} \cite{cfg}.
\par
We  can use the previous observations to construct U-invariants 
for some  $N=2$ special geometries looking at the representation content of
 vectors and their duals
with respect to U-dualities.

Let us first consider $N=2$ theories with U-duality $SU(1,n)$ and $SU(3,3)$.
These groups emerge in discussing string compactifications on some $N=2$
 orbifolds
(i.e. orbifold points of Calabi-Yau threefolds)\cite{cfg}\cite{seib}.

The vector content is respectively given by the fundamental representation of
 $SU(1,n)$
and the twenty dimentional threefold antisymmetric rep. of $SU(3,3)$
 \cite{ffs}.

Amazingly, the first representation occurs as in $N=3$ matter coupled theories,
while the latter  is the same as in $N=5$ supergravity
(note that $SU(1,n)$,  $SU(3,n)$ and  $SU(3,3)$,  $SU(5,1)$ are just different
non compact forms of the same $SU(m)$ groups).

From the results of the previous section we conclude that
  the special manifolds  ${SU(1,n)\over SU(n)\times U(1)}$
and ${SU(3,3)\over SU(3)\times SU(3)\times U(1)}$  
 admit respectively a quadratic \cite{cedafe}, \cite{sabra} and
a quartic topological invariant.
The $N=2$  special manifold ${O^*(12)\over U(6)}$ has a vector content 
which is a left spinor of $O^*(12)$, as in the $N=6$ theory, therefore it
 admits 
a quartic invariant.

Finally, the $N=2$ special manifolds ${SU(1,1) \over U(1)}
 \times{O(2,n)\over O(2) \times O(n)}$, which emerge in $N=2$ 
compactifications 
of both heterotic and Type II strings \cite{seib}, admit a quartic invariant
 which can be read
from  the $N=4$ quartic invariant in which the  ${SU(1,1) \over U(1)}$ matter 
charge is identified 
with the second eigenvalue of the $N=4$ central charge.

All the above topological invariants can then be interpreted as entropy 
of a variety of $N=2$ black-holes.


\section*{Appendix: A simple determination of the U-invariants }
 In order to determine the quartic U-invariant expressions $S^2$ , $\nabla S
 =0$, of the $N>4$ theories, it is useful to use, as a calculational tool, 
transformations
 of the coset which preserve the normal form of the $Z_{AB}$ matrix.
 It turns out that these transformations are certain Cartan elements
 in $K=G/H$ \cite{solv}, that is they belong to $O(1,1)^p\in K$, with $p=1$ 
for $N=5$,
 $p=3$ for $N=6,8$.
 \par
 These elements act only on the $Z_{AB}$ (in normal form), but they
 uniquely determine the U-invariants since they mix the eigenvalues
 $e_i$ ($i=1,\cdots,[N/2]$).
 \par
 For $N=5$, $SU(5,1)/U(5)$ has rank one (see ref. \cite{gilmore}) and
 the element is:
 \begin{equation}
\delta e_1 = \xi e_2 ; \quad \delta e_2 = \xi e_1
\end{equation}
which is indeed a $O(1,1)$ transformation with unique invariant
\begin{equation}
\vert (e_1)^2 - (e_2) ^2\vert = {1 \over 2} \sqrt{ 8 \left((e_1)^4 + (e_2)^4
\right) - 4 \left( (e_1)^2 + (e_2)^2 \right)^2 }
\end{equation}

For $N=6$, we have $\xi_1 \equiv \xi_{12}; \xi_2 \equiv \xi_{34};
\xi_3 \equiv \xi_{56}$ and we obtain the 3 Cartan elements of
$O^*(12) /U(6)$, which has rank 3, that is it is a $O(1,1)^3$ in $O^*
(12)/U(6)$.
Denoting by $e$ the singlet charge, we have the following $O(1,1)^3$
transformations:
\begin{eqnarray}
\delta e_1 &=& \xi_2 e_3  + \xi _3 e_2   + \xi _1 e \label{trans61}\\
 \delta e_2 &=& \xi_1 e_3  + \xi _3 e_1   + \xi _2 e\label{trans62} \\
 \delta e_3 &=& \xi_1 e_2  + \xi _2 e_1   + \xi _3 e\label{trans63} \\
 \delta e &=& \xi_1 e_1  + \xi _2 e_2   + \xi _3 e_3\label{trans64}
\end{eqnarray}
these transformations fix uniquely the $O^*(12)$ invariant constructed out of
the five $U(6)$ invariants displayed in (\ref{invar61}-\ref{invar65}).

For $N=8$ the infinitesimal parameter is $\xi_{ABCD}$ and, using the
reality condition, we get again a $O(1,1)^3$ in $E_{7(-7)} /SU(8)$.
Setting $\xi_{1234}= \xi_{5678} \equiv \xi_{12}$, $\xi_{1256}= 
\xi_{3478} \equiv \xi_{13}$,
$\xi_{1278}= \xi_{3456} \equiv \xi_{14}$, we have the following set of
transformations:
\begin{eqnarray}
 \delta e_1 &=& \xi_{12} e_2  + \xi _{13} e_3   + \xi _{14} e_4\label{trans81}
 \\
 \delta e_2 &=& \xi_{12} e_1  + \xi _{13} e_4   + \xi _{14} e_3\label{trans82}
 \\
 \delta e_3 &=& \xi_{12} e_4  + \xi _{13} e_1   + \xi _{14} e_2\label{trans83}
 \\
 \delta e_4 &=& \xi_{12} e_3  + \xi _{13} e_2   + \xi _{14} e_1\label{trans84}
\end{eqnarray}

These transformations fix uniquely the relative coefficients of the three
 $SU(8)$
invariants:
\begin{eqnarray}
I_1&=& e_1^4 + e_2^4 + e_3^4 + e _4^4 \\
I_2&=& (e_1^2 + e_2^2 + e_3^2 + e _4^2)^2 \\
I_3&=& e_1e_2e_3e _4 \\
\end{eqnarray}

It is easy to see that the transformations (\ref{trans61}-\ref{trans64}) and
 (\ref{trans81}-\ref{trans84}) correspond to three commuting matrices 
(with square equal 
to $\bfone$):
\begin{equation}
\pmatrix{0&0&0&1 \cr 0&0&1&0 \cr 0&1&0&0 \cr 1&0&0&0 \cr} ; 
\pmatrix{0&1&0&0 \cr 1&0&0&0 \cr 0&0&0&1 \cr 0&0&1&0 \cr} ;
\pmatrix{0&0&1&0 \cr 0&0&0&1 \cr 1&0&0&0 \cr 0&1&0&0 \cr}
\end{equation}
which are proper non compact Cartan elements of $K$.
The reason we get the same transformations for $N=6$ and $N=8$
is because the extra singlet $e$ of $N=6$ can be identified 
with the fourth eigenvalue of the central charge of $N=8$.


\begin{thebibliography}{99}
\bibitem{string}
For a review, see for instance:
M. J. Duff, R. R. Khuri and J. X. Lu, {\it String solitons}, Phys. Rep.
 {\bf 259} (1995) 213;
M. J. Duff, Kaluza-Klein theory in perspective, in {\it Proceedings of
 the Nobel Symposium Oskar
Klein Centenary, Stockholm, September 1994} (World Scientific, 1995),
 E. Lindstrom editor, hep-th/9410046;
G. Horowitz, UCSBTH-96-07, gr-qc/9604051;
J. M. Maldacena, Ph.D. thesis, hep-th/9607235;
M. Cvetic, UPR-714-T, hep-th/9701152
\bibitem{black}
G. Gibbons, in {\it Unified theories of Elementary Particles. Critical 
Assessment
and Prospects}, Proceedings of the Heisemberg Symposium, M\"unchen, West
 Germany, 1981,
 ed. by P. Breitenlohner and H. P. D\"urr, Lecture Notes in Physics Vol.
 160 (Springer-Verlag, Berlin, 1982);
G. W. Gibbons and C. M. Hull, Phys. lett. {\bf 109B} (1982) 190;
G. W. Gibbons, in {\it Supersymmetry, Supergravity and Related Topics},
 Proceedings
of the XVth GIFT International Physics, Girona, Spain, 1984, ed. by F. del
 Aguila, J.
de Azc\'arraga and L. Ib\'a\~nez, (World Scientific, 1995), pag. 147;
R. Kallosh, A. Linde, T. Ortin, A. Peet and A. Van Proeyen, Phys. Rev.
 {\bf D46} (1992) 5278;
R. Kallosh, T. Ortin and A. Peet, Phys. Rev. {\bf D47} (1993) 5400;
R. Kallosh, Phys. Lett. {\bf B282} (1992) 80;
R. Kallosh and A. Peet, Phys. Rev. {\bf D46} (1992) 5223;
A. Sen, Nucl. Phys. {\bf B440} (1995) 421; Phys. Lett. {\bf B303} (1993) 221;
 Mod. Phys. Lett. {\bf A10}
(1995) 2081;
J. Schwarz and A. Sen, Phys. Lett. {\bf B312} (1993) 105;
M. Cvetic and D. Youm, Phys. Rev. {\bf D53} (1996) 584;
M. Cvetic and A. A. Tseytlin, Phys. Rev. {\bf D53} (1996) 5619;
M. Cvetic and C. M. Hull, Nucl. Phys. {\bf B480} (1996) 296
\bibitem{dbr}
A. Strominger and C. Vafa, Phys. Lett. {\bf B379} (1996) 99, hep-th/9601029;
C. G. Callan and J. M. Maldacena, Nucl. Phys. {\bf B472}
 (1996) 591, hep-th/9602043; 
G. Horowitz and A. Strominger, Phys. Rev. Lett. {\bf B383} (1996) 2368, 
 hep-th/9602051;
R. Dijkgraaf, E. Verlinde, H. Verlinde, Nucl.Phys. {\bf B486} (1997) 77,
 hep-th/9603126; 
P. M. Kaplan, D. A. Lowe, J. M. Maldacena and A. Strominger, hep-th/9609204;
 J. M. Maldacena, hep-th/9611163
\bibitem{micros}
L. Susskind, hep-th/9309145; L. Susskind and J. Uglum, Phys. Rev. {\bf D50}
 (1994) 2700;
F. Larsen and F. Wilczek, Phys. Lett. B375 (1996) 37, hep-th/9511064
\bibitem{fks}
S. Ferrara, R. Kallosh and A. Strominger, Phys. Rev. {\bf D52} (1995) 5412,
 hep-th/9508072;
A. Strominger, Phys. Lett. {\bf B383} (1996) 39, hep-th/9602111
\bibitem{feka1}
S. Ferrara and R. Kallosh, Phys. Rev. {\bf D54} (1996) 1514, hep-th/9602136
\bibitem{feka2}
S. Ferrara and R. Kallosh, Phys. Rev. {\bf D54} (1996) 1525, hep-th/9603090
\bibitem{fegika}
S. Ferrara, G. W. Gibbons and R. Kallosh, hep-th/9702103
\bibitem{kall}
R. Kallosh and B. Kol, Phys. Rev. {\bf D53} (1996) 5344
\bibitem{bmgk}
P. Breitenlohner, D. Maison and G. W. Gibbons, Commun. Math. Phys. {\bf 120}
 (1988) 295;
G. W. Gibbons, R. Kallosh and B. Kol, Phys. Rev. Lett. {\bf 77} (1996) 4992,
 hep-th/9607108 
\bibitem{fesazu}
S. Ferrara, C. Savoy and B. Zumino, Phys. Lett. {\bf 100B} (1981) 393
\bibitem{cedafe}
A. Ceresole,  R. D'Auria and S. Ferrara, in ``{\it S-Duality and Mirror
 symmetry}'',
Nucl. Phys. (Proc. Suppl.) {\bf B46} (1996) 67, ed. E. Gava, K. S. Narain 
and C. Vafa, 
hep-th/9509160
\bibitem{noi1}
L. Andrianopoli, R. D'Auria and S. Ferrara, hep-th/9608015, to appear in
 International Journal of Modern Physics A
\bibitem{noi}
L. Andrianopoli, R. D'Auria and S. Ferrara, hep-th/9612105
\bibitem{cj}
E. Cremmer in ``{\it Supergravity '81}'', ed. by S. Ferrara and J. G. Taylor,
 Pag. 313;
B. Julia in ``{\it Superspace \& Supergravity}'', ed. by S. Hawking and M.
 Rocek, Cambridge (1981) pag. 331
\bibitem{maina}
L. Castellani, A. Ceresole, R. D'Auria, S. Ferrara, P. Fr\'e and E. Maina,
 Nucl. Phys. {\bf B286} (1986)
317
\bibitem{bks}
E. Bergshoeff, I. G. Koh and E. Sezgin, Phys. Lett. {\bf 155B} (1985) 71;
M. de Roo and F. Wagemans, nucl. Phys. {\bf B262} (1985) 644
\bibitem{ht}
C. M. Hull and P. K. Townsend, Nucl. Phys. {\bf B451} (1995) 525,
 hep-th/9505073
\bibitem{witten}
E. Witten, ``{\it String Theory Dynamics in Various Dimensions}'', Nucl. Phys.
 {\bf B433},
hep-th/9503124
\bibitem{dlr}
M. J. Duff, J. T. Liu and J. Rahmfeld, ``{\it Four dimensional
 string/string/string triality}'',
Nucl. Phys. {\bf B459} (1996) 125
\bibitem{fkff}
S. Ferrara and C. Kounnas, Nucl. Phys. {\bf B328} (1989) 406;
S. Ferrara and P. Fr\'e, Int. Jour. Mod. Phys. A {\bf Vol. 5, No. 5}
 (1990) 989
\bibitem{crvp}
E. Cremmer and A. Van Proeyen, Class. Quant. Grav. {\bf 2} (1985) 445
\bibitem{cfg}
S. Cecotti, S. Ferrara and L. Girardello, Int. Jour. Mod. Phys. A
 {\bf Vol. 4} (1989) 2475
\bibitem{seib}
N. Seiberg, Nucl. Phys. {\bf B303} (1988) 286;
L. Dixon, V. Kaplunowsky, J. Louis, Nucl. Phys. {\bf B329} (1990) 27;
S. Ferrara, C. Kounnas and  M. Porrati, Phys. Lett. {\bf B181} (1986) 26
\bibitem{ffs}
S. Ferrara, P. Fr\'e and P. Soriani, Class. Quantum Grav. {\bf 9} (1992) 1649
\bibitem{sabra}
K. Behrndt, W. A. Sabra, hep-th/9702010 
\bibitem{solv}
L. Andrianopoli, R. D'Auria, S. Ferrara, P. Fr\'e and M. Trigiante,
  hep-th/9611014 
\bibitem{gilmore}
R. Gilmore, ``{\it Lie groups, Lie algebras and some of their applications}'',
 (1974) ed. J. Wiley and sons      
\end{thebibliography}
\end{document}